\documentclass{article}
\usepackage{spconf,amsmath,graphicx}
\usepackage{epsfig}
\usepackage{xspace}
\usepackage{srcltx}
\usepackage{caption}
\usepackage{subcaption}
\usepackage[export]{adjustbox}
\usepackage{multirow,makecell,hhline}
\usepackage{booktabs}
\usepackage{textcomp}
\usepackage{cite}
\usepackage{hyperref}
\usepackage{CJKutf8}

\usepackage{color}
\usepackage[dvipsnames]{xcolor}

\newcommand{\CMT}[1]{{}}

\newcommand{\tbh}[1]{\textbf{#1}}

\title{Bytes are All You Need:\\ End-to-End Multilingual Speech Recognition and Synthesis with Bytes}
\name{Bo Li, Yu Zhang, Tara  Sainath, Yonghui Wu, William Chan}
\address{Google \\
\texttt{
\{boboli,ngyuzh,tsainath,yonghui,williamchan\}@google.com}}

\begin{document}
\ninept
\maketitle
\begin{abstract}
We present two end-to-end models: Audio-to-Byte (A2B) and Byte-to-Audio (B2A), for multilingual speech recognition and synthesis. Prior work has predominantly used characters, sub-words or words as the unit of choice to model text. These units are difficult to scale to languages with large vocabularies, particularly in the case of multilingual processing. In this work, we model text via a sequence of Unicode bytes, specifically, the UTF-8 variable length byte sequence for each character. Bytes allow us to avoid large softmaxes in languages with large vocabularies, and share representations in multilingual models. We show that bytes are superior to grapheme characters over a wide variety of languages in monolingual end-to-end speech recognition. Additionally, our multilingual byte model outperform each respective single language baseline on average by 4.4\% relatively. In Japanese-English code-switching speech, our multilingual byte model outperform our monolingual baseline by 38.6\% relatively. Finally, we present an end-to-end multilingual speech synthesis model using byte representations which matches the performance of our monolingual baselines.
\end{abstract}


\vspace{-0.05in}
\section{Introduction \label{sec:introduction}}
\vspace{-0.05in}

Expanding the coverage of the world's languages in Automatic Speech Recognition (ASR) and Text-to-Speech (TTS) systems have been attracting much interest in both academia and industry \cite{schultz2006multilingual, bourlard2011current}. Conventional phonetically-based speech processing systems require pronunciation dictionaries that map phonetic units to words. Building such resources require expert knowledge for each language. Even with the costly human effort involved, many languages do not have sufficient linguistic resources available for building such dictionaries. Additionally, the inconsistency in the phonetic systems is also challenging to resolve \cite{gales2015unicode} when merging different languages.

Graphemes have been used as an alternative modeling unit to phonemes for speech processing \cite{kanthak2002context,killer2003grapheme,stuker2004grapheme,basson2012comparing}. For these systems, an orthographic lexicon instead of a pronunciation dictionary is used to provide a vocabulary list. With recent advances in end-to-end (E2E) modeling, graphemes have become a popular choice. For example, \cite{graves2014towards} built a Connectionist Temporal Classification (CTC) model to directly output graphemes, while \cite{ChanJaitlyLeEtAl15,bahdanau2016end,chan-interspeech-2016} used graphemes in sequence-to-sequence (seq2seq) models. Sub-word units were used in seq2seq \cite{chan-iclr-2017,cc-icassp-2018,zeyer-interspeech-2018} and RNNT \cite{Rao2017} models, and word units were used by \cite{soltau-arxiv-2016,li-icassp-2018}. Similarly, graphemes are also commonly used to build end-to-end TTS systems \cite{sotelo2017char2wav,wang2017tacotron,shen2018natural}.

The use of graphemes bring model simplicity and enables end-to-end optimization, which has been shown to yield better performance than phoneme-based models \cite{prabhavalkar2017comparison}. However, unlike phonemes, the size of the grapheme vocabulary varies greatly across languages. For example, many eastern languages, such as Chinese, Japanese and Korean, have tens of thousands of graphemes. With limited amounts of training data, many graphemes may have little or no coverage. The label sparsity issue becomes even more severe for multilingual models, where one needs to pool all the distinct graphemes from all languages together resulting in a very large vocabulary that often has long tail graphemes with very poor coverage.

To address these problems, \cite{gales2015unicode} explored the use of features from Unicode character descriptions to construct decision trees for clustering graphemes. However, when the model changes to support a new language, the decision tree needs to be updated. Recently, there has been work on exploring the use of Unicode bytes to represent text. \cite{gillick2015multilingual} presented an LSTM-based multilingual byte-to-span model. The model consumes the input text byte-by-byte and outputs span annotations. The Unicode bytes are language independent and hence a single model can be used for many languages. The vocabulary size of Unicode bytes is always 256 and it does not increase when pooling more languages together, which is more preferable to graphemes for multilingual applications. 

In this work, we investigate the potential of representing text using byte sequences introduced in \cite{gillick2015multilingual} for speech processing. For ASR, we adopt the Listen, Attend and Spell (LAS)\cite{ChanJaitlyLeEtAl15} model to convert input speech into sequences of Unicode bytes which correspond to the UTF-8 encoding of the target texts. This model is referred to as the Audio-to-Byte (A2B) model. For TTS, our model is based on the Tacotron 2 architecture \cite{shen2018natural}, and generates speech signals from an input byte sequence. This model is referred to as the the Byte-To-Audio (B2A) model. Since both the A2B model and the B2A model operate directly on Unicode bytes, they can handle any number of languages written in Unicode without any modification to the input processing. Due to the small vocabulary size being used, 256 in this case, our models can be very compact and very suitable for on-device applications.

We report recognition results for the A2B model on 4 different languages -- English, Japanese, Spanish and Korean. First, for each individual language, we compare our A2B models to Audio-to-Char (A2C) models which emit grapheme outputs. For English and Spanish where the graphemes are single-byte characters, A2B has the exact same performance as A2C as expected. However, for languages that have a large grapheme vocabulary, such as Japanese and Korean, the label sparsity issue hurts the performance of A2C models, whereas the A2B model shares bytes across graphemes and performs better than A2C models.
Benefiting from the language independence representation of Unicode bytes, we find it is possible to progressively add support for new languages when building a multilingual A2B model.
Specifically, we start with an A2B model trained on English and Japanese and add in a new language after convergence. When adding a new language we usually make sure the new language has the highest mixing ratio but meanwhile keeping small portion for each of the existing languages to avoid forgetting older ones. We experiment adding Spanish and Korean one at a time. In this way, we can reuse the previously built model and expand the language coverage without modifying the model structure. For multilingual ASR, we find that the A2B trained in this way is better than training from scratch. In addition, by adding a 1-hot language vector to the A2B system, which has been shown to boost multi-dialect \cite{li2018multi} and multilingual \cite{toshniwal2018multilingual} system performance, we find that the multilingual A2B system outperforms all the language dependent ones. 

We evaluate the B2A model on 3 different languages, which include English, Mandarin and Spanish. Again, we compare B2A models with those take graphemes as input. For all three languages, B2A has similar performance on quantitative subjective evaluations as graphemes trained on single languages, this providing a more compact multilingual TTS model.

\section{Multilingual Audio-to-Byte (A2B)}
\label{sec:asr}
\vspace{-0.05in}

\subsection{Model Structure}
\vspace{-0.05in}
The Audio-to-Byte (A2B) model is based on the Listen, Attend and Spell (LAS) \cite{ChanJaitlyLeEtAl15} model, with the output target changed from graphemes to Unicode bytes. The encoder network consists of 5 unidirectional Long Short-Term Memory (LSTMs) \cite{hochreiter1997long} layers, with each layer having $1,400$ hidden units. The decoder network consists of 2 unidirectional LSTM layers with $1,024$ hidden units. Additive content-based attention \cite{bahdanau2014neural} with 4 attention heads are used to learn the alignment between the input audio features and the output target units. The output layer is a 256 dimensional softmax, corresponding to the 256 possible byte values.

Our front-end consists of 80-dimensional log-mel features, computed with a 25ms window and shifted every 10ms. Similar to \cite{pundak2016lower,sak2015fast}, at each current frame, these features are stacked with 3 consecutive frames to the left and then down-sampled to a 30ms frame rate. 
The amount of training data usually varies across languages. For example, for English we have around 3.5 times the amount of data compared to the other languages. More details about data can be found in Section~\ref{sec:results}. In this work, we adjust the data sampling ratio of the different languages to help tackle the data imbalance. We choose the sampling ratio based on intuition and empirical observations. Specially, we start with mixing the language equally and increase the ratio for a language where the performance needs more improvement. 
In addition, a simple 1-hot language ID vector has been found to be effective improving multilingual systems \cite{li2018multi,toshniwal2018multilingual}. We also adopt this 1-hot language ID vector as additional input passed into the A2B models, and concatenate it to all the layers including both the encoder and decoder layers.

\subsection{Output Unit}
\vspace{-0.05in}
End-to-end speech recognition models have typically used characters \cite{ChanJaitlyLeEtAl15}, sub-words \cite{chan-iclr-2017}, word-pieces \cite{Rao2017} or words \cite{soltau-arxiv-2016} as the output unit of choice. Word-based units are difficult to scale for languages with large vocabularies, which makes the softmax prohibitively large, especially in multilingual models. One solution is to use data-driven word-piece models. Word-pieces learned from data can be trained to have a fixed vocabulary size. But it requires building a new word-piece model when a new language or new data is added. Additionally, building a multilingual word-piece model is challenging due to the unbalanced grapheme distribution. Grapheme units give the smallest vocabulary size among these units; however,  some languages still have very large vocabularies. For example our Japanese vocabulary has over $4.8$k characters. In this work, we explore decomposing graphemes into a sequence of Unicode bytes.

Our A2B model generates the text sequence one Unicode byte at a time. We represent text as a sequence of variable length UTF-8 bytes. For languages with single-byte characters (e.g., English), the use of byte output is equivalent to the grapheme character output. However, for languages with multi-byte characters, such as Japanese and Korean, the A2B model needs to generate a sequence of correct bytes to emit one grapheme token. This requires the model to learn both the short-term within-grapheme byte dependencies, and the long-term inter-grapheme or even inter-word/phrase dependencies, which would be a harder task than grapheme based system.

The main advantage of byte representation is its language independence. Any script of any language representable by Unicode can be represented by a byte sequence, and there is no need to change the existing model structure. However, for grapheme models, whenever there is a new symbol added, there is a need to change the output softmax layer. This language independence makes it more preferable for modeling multiple languages and also code-switching \cite{auer2013code} speech within a single model.
\section{Multilingual Byte-to-Audio (B2A)}
\label{sec:tts}
\vspace{-0.05in}

\subsection{Model Structure}
\vspace{-0.05in}
The Byte-to-Audio (B2A) model is based on Tacotron 2\cite{shen2018natural} model. The input byte sequence embedding is encoded by three convolutional layers, which contain 512 filters with shape $5 \times 1$, followed by a bidirectional long short-term memory (LSTM) layer of 256 units for each direction. The resulting text encodings are accessed by the decoder through a location sensitive attention mechanism, which takes attention history into account when computing a normalized weight vector for aggregation.

The autoregressive decoder network takes as input the aggregated byte encoding, and conditioned on a fixed speaker embedding for each speaker, which is essentially the language ID since our training data has only one speaker per language.
Similar to Tacotron 2, we separately train a WaveRNN~\cite{kalchbrenner2018efficient} to invert mel spectrograms to a time-domain waveform.

\section{Results}
\label{sec:results}
\vspace{-0.05in}

\begin{table*}[t]
\caption{Speech recognition performance of monolingual and multilingual with Audio-to-Byte (A2B) or Audio-to-Char (A2C) models.}
\vspace{-0.1in}
\centering
\begin{tabular}{c|c|l|l||c|c|c|c}
\toprule
\multirow{2}{*}{\tbh{Model}} & \multirow{2}{*}{\tbh{ExpId}} & \multirow{2}{*}{\tbh{Configuration}} & \tbh{Training} & \tbh{English} & \tbh{Japanese} & \tbh{Spanish} & \tbh{Korean} \\
~ & ~ & ~ & \tbh{Languages} & \tbh{WER(\%)} & \tbh{TER(\%)} & \tbh{WER(\%)} & \tbh{WER(\%)} \\
\midrule
\multirowcell{2}{\tbh{Mono-}\\\tbh{lingual}} & A1 & A2C & \multirow{2}{*}{EN/JA/ES/KO} & 6.9 & 13.8 & 11.2 & 26.5 \\
~ & A2 & A2B & ~ & 6.9 & {\bf 13.2} & 11.2 & {\bf 25.8} \\
\midrule
\multirowcell{11}{\tbh{Multi-}\\\tbh{lingual}} & B1 & A2C & \multirow{2}{*}{EN+JA} & 9.5 & 13.9 & - & - \\
~ & B2 & A2B & ~ & 8.9 & {\bf 13.3} & - & -\\
\cline{2-8}
~ & C1 & A2B, Random Init & \multirow{2}{*}{EN+JA+ES}  & 9.7 & 13.6 & 11.1 & -\\
~ & C2 & A2B, Init From B2 & ~ & 8.6 & {\bf 13.2} & {\bf 11.0} & - \\
\cline{2-8}
~ & D1 & A2B, Init From C2  & EN+JA+ES+KO & 8.4 & 13.4 & 11.3 & 26.0 \\
\hhline{~=======}
~ & B3 & A2B, Larger Model & \multirow{2}{*}{EN+JA} & 8.8 & 13.6 & - & - \\
~ & B4 & A2B, Larger Model, LangVec & ~ & 7.5 & 13.3 & - & - \\
\cline{2-8}
~ & C3 & A2B, Init From B4 & EN+JA+ES & 7.5 & 12.9 & 10.8 & - \\
\cline{2-8}
~ & D2 & A2B, Larger Model, LangVec & \multirow{3}{*}{EN+JA+ES+KO} & 8.6 & 13.5 & 11.2 & 25.4 \\
~ & D3 & A2B, Init From C3 & ~ & 7.0 & 12.8 & 10.8 & 25.0 \\
~ & D4 & A2B, Init From D3 & ~ & {\bf 6.6} & {\bf 12.6} & {\bf 10.7} & {\bf 24.7} \\
\bottomrule
\end{tabular}
\label{tbl:multilang_asr}
\vspace{-0.1in}
\end{table*}

\subsection{Byte for ASR}
\vspace{-0.05in}

\begin{table}[t]
\caption{Statistics of the training and testing data used in our experiments. ``utts'' denotes the total number of utterances in each set and ``time'' is the total duration of audio for each set.}
\vspace{-0.1in}
\centering
\begin{tabular}{c||r|r|r|r}
\toprule
\multirow{2}{*}{\tbh{Languages}} & \multicolumn{2}{c|}{\tbh{Train}}  & \multicolumn{2}{c}{\tbh{Test}}  \\
\cline{2-5}
~ & \tbh{utts (M)} & \tbh{time (Kh)} & \tbh{utts (K)} & \tbh{time (h)} \\
\midrule
English (EN) & 35.0 & 27.5 & 15.4 & 20.0 \\
Japanese (JA) & 9.9 & 16.5 & 17.6 & 22.2 \\
Spanish (ES) & 8.9 & 16.3 & 16.6 & 22.3 \\
Korean (KO) & 9.6 & 16.1 & 12.6 & 15.0\\
\bottomrule
\end{tabular}
\label{tbl:data}
\vspace{-0.2in}
\end{table}

\subsubsection{Data}
\vspace{-0.05in}

Our speech recognition experiments are conducted on a human transcribed supervised training set consisting speech from 4 different languages, namely English (EN), Japanese (JA), Spanish (ES) and Korean (KO). The total amount of data is around 76,000 hours and the language-specific information can be found in Table~\ref{tbl:data}. These training utterances are anonymized and hand-transcribed, and are representative of Google’s voice search and dictation traffic. These utterances are further artificially corrupted using a room simulator \cite{Chanwoo17}, adding varying degrees of noise and reverberation such that the overall SNR is between 0dB and 30dB, with an average SNR of 12dB. The noise sources are from YouTube and daily life noisy environmental recordings. For each utterance, we generated 10 different noisy versions for training.
For evaluation, we report results on language-specific test sets, each contains roughly 15K anonymized, hand-transcribed utterances from Google’s voice search traffic without overlapping with the training data. This amounts to roughly 20 hours of test data per language. Details of each language dependent test set can be found in Table~\ref{tbl:data}. We use word error rates (WERs) as the evaluation criterion for all the languages except for Japanese, where token error rates (TERs) are used to exclude the ambiguity of word segmentation.

\subsubsection{Language Dependent Systems}
\vspace{-0.05in}

We first build language dependent A2B models to investigate the performance of byte-based language representations for ASR. For comparison, we also build corresponding Audio-to-Char (A2C) models that have the same model structure but output graphemes. For all the four languages, the model which outputs byte always has a 256-dimensional softmax output layer. However, for the grapheme models, different grapheme vocabularies have to be used for different languages. The grapheme set is complete for English and Spanish as it contains all possible letters in each of the languages. However, for Japanese and Korean, we use the training data vocabularies which are $4.8$K and $2.7$K respectively. The corresponding test set grapheme OOV rates are 2.1\% and 1.0\%. Whereas with byte outputs, we do not have OOV problem for any language.

Experimental results are presented as {\tt A1} for the A2C models and {\tt A2} for the A2B models in Table~\ref{tbl:multilang_asr}. The difference between grapheme and byte representations mainly lies in languages which use multi-byte characters, such as Japanese and Korean. Comparing {\tt A1} to {\tt A2}, byte outputs give better results for Japanese and Korean. While for languages with single-byte characters, namely English and Spanish, they have exactly the same performance as expected. Byte output requires the model to learn both the short-term within-grapheme byte dependencies and the long-term inter-grapheme or even inter-word/phrase dependencies; it would possibly be a harder task than grapheme based systems. However, the A2B model yields a 4.0\% relative WER reduction on Japanese and 2.6\% on Korean over the grapheme systems. It is interesting to see that even with the same model structure, we are able to get better performance with the byte representation.

\subsubsection{Multilingual ASR Systems}
\vspace{-0.05in}

In this experiment, we justify the effectiveness of byte based models over graphemes for multilingual speech recognition. We first build a joint English and Japanese model by equally mixing the training data. For grapheme system, we combine the grapheme vocab of English and Japanese which leads to a large softmax layer. The same model structure except for the softmax layer, where a 256 dimensional softmax is used, is used to build the A2B model. Although the model now needs to recognize two languages, we keep the model size the same as those language dependent ones. From Table~\ref{tbl:multilang_asr}, the multilingual byte system ({\tt B2}) is better than the grapheme system ({\tt B1}) on both English and Japanese test sets. However, its performance is worse than those language dependent ones,
which we will address later in this work. For the following experiments, we continue with only the A2B models as they are better than A2C models.

To increase the model's language coverage, e.g., Spanish, one way is to start from a random initialization and train on all the training data. We equally mix the data from these three languages for training. The results are presented as {\tt C1} in Table~\ref{tbl:multilang_asr}. Due to the language independence of the byte representation, we, alternatively, can add a new language by simply training on new data. Hence, we reuse the {\tt B2} model to continue training with Spanish data. To avoid the model forgetting previous languages, namely English and Japanese, we also mix in those languages but with a slightly lower mixing ratio which is 3:3:4 for English, Japanese and Spanish. The results are presented as {\tt C2} in Table~\ref{tbl:multilang_asr}. With this method, the byte model not only trains faster but also achieves better performance than {\tt C1}. Most importantly, {\tt C2} matches the performance of language dependent models on Japanese and is even slightly better for Spanish.

To add support for Korean, we simply continued the training of {\tt C2} with the new training data mixture. We use a ratio of 0.23:0.23:0.23:0.31, which is based on heuristics to balance the existing languages and use a higher ratio for the new languages.
We did not specifically tune the mixing ratio. The results ({\tt D1} in Table~\ref{tbl:multilang_asr}) show that we are able to get closer to the language dependent models except for English. Even though worse than the English only model, {\tt D1} gives the best multilingual performance on English so far.

To improve the performance of the multilingual systems, we first increase the number of decoder layers from 2 to 6 in consideration of the increased variations in byte sequences when mixing more languages. However, experimental results show that the larger model improves performance on English but degrades on Japanese due to potential over-fitting (comparing {\tt B3} to {\tt B2}). To address this problem, we brings in the 1-hot language ID vector to all the layers in the A2B model. This enables the learning of language independent weight matrices together with language dependent biases to cater the specific needs for each language. Experiment {\tt B4} shows dramatic error reduction with this simple 1-hot vector comparing to {\tt B3}.

Similarly, to support the recognition of Spanish, we continue the training of {\tt B4} by mixing the languages at the ratio of 3:3:4 where more weight is given to the new language. This gives us the model {\tt C3} which outperforms language dependent ones on both Japanese and Spanish.
Furthermore, we add Korean in a similar way with the ratio of 0.3:0.15:0.15:0.4. This time while making sure the ratio for the new language, Korean, is the highest, we also increase the ratio for English as we have more English training data. The model {\tt D3} wins over language dependent models except for English. One assumption for the degradation on English is that when mixing in other languages, the multilingual model sees less data from each language than those single language models. To justify this, we continue the training of {\tt D3} with an increased English data presence ratio in the mixture, specifically we use the ratio of 2:1:1:1. We didn't specifically tune these mixing ratios used for training.
The final model {\tt D4} wins over all the language dependent ones on average by 4.4\% relatively. For comparison, we include the results for a randomly initialized model with equal training data mixing ratio {\tt D2}, which is much worse.

\subsubsection{Error analysis}
\vspace{-0.05in}

\begin{table}[t]
\caption{Results on A2B and A2C models on English-Japanese code-switching data.}
\vspace{-0.1in}
\centering
\begin{tabular}{c|c|l||c}
\toprule
\tbh{Model} & \tbh{ExpId} & \tbh{Configuration} & \tbh{TER(\%)} \\
\midrule
\multirowcell{2}{\tbh{Mono-}\\\tbh{lingual}} & A1 & A2C & 36.5 \\
~ & A2 & A2B & 22.4 \\
\midrule
\multirowcell{3}{\tbh{Multi-}\\\tbh{lingual}} & B1 & A2C & 21.4 \\
~ & B2 & A2B & {\bf 20.5} \\
\cline{2-4}
~ & D4 & A2B Larger Model, LangVec & 21.3 \\
\bottomrule
\end{tabular}
\label{tbl:codeswitch}
\vspace{-0.2in}
\end{table}

To further understand the gains of using bytes versus graphemes as language representations, we take Japanese for this study and compare the decoding hypotheses between {\tt A1} and {\tt A2}. Interestingly, the A2B model wins over the A2C models mainly on English words in utterances with mixed English and Japanese. The Japanese test set was not particularly created to include code-switching utterances. Examining the English words appeared in Japanese test set, they are mostly proper nouns such as ``{\it Google}'', ``{\it wi-fi}'', ``{\it LAN}'' etc. One example of such cases is the A2B generates the correct hypothesis ``{\it wi-fi \begin{CJK}{UTF8}{min}オ ン\end{CJK}}'' while the A2C outputs ``{\it i-i \begin{CJK}{UTF8}{min}オ ン\end{CJK}}''. Another example is ``{\it google \begin{CJK}{UTF8}{min}音 声 認 識\end{CJK}}'' where the A2B recognizes it correctly, but the A2C model drops the initial ``{\it g}'' and gives ``{\it oogle \begin{CJK}{UTF8}{min}音 声 認 識\end{CJK}}''.

One of the potential benefits of using byte-based models is for code-switching speech. Collecting such data is challenging. The quality of artificially concatenated speech is far from real. In this study we use data filtered from the Japanese test set, where utterances having transcript that contains 5 or more consecutive English characters are kept. These utterances  mostly contain only a single English word in Japanese texts. Out of the 17.6K utterances, we get 476 code-switching sentences and we report the TERs on this subset in Table~\ref{tbl:codeswitch}. With Japanese monolingual models ({\tt A1} and {\tt A2}), our A2B model outperforms the A2C model by 38.6\% relatively. With English and Japanese multilingual models ({\tt B1} and {\tt B2}), our A2B model wins over the A2C model by 4.2\% relatively. We also test system {\tt D4} on these code-switch data. However, due to the language 1-hot vector used in {\tt D4} is utterance-level, the performance is worse than {\tt B2}. Using frame/segment level language information may address this problem, which will be explored in future. 

\subsection{Byte for TTS}
\vspace{-0.05in}

\subsubsection{Data}
\vspace{-0.05in}
Text-to-speech models were trained on (1) 44 hours
of North American English speech recorded by a female
speaker; (2) 37 hours of Mandarin speech by a female speaker; (3) 44 hours of North American Spanish speech by a female speaker. For all compared models, we synthesize
raw audio at 24 kHz in 16-bit format. We rely on crowdsourced Mean Opinion Score (MOS) evaluations based on subjective listening tests. 
All our MOS evaluations are aligned to the \textit{Absolute Category Rating} scale~\cite{rec1996p}, with rating scores from 1 to 5 in 0.5~point increments. 

\subsubsection{Multilingual TTS System}
\vspace{-0.05in}
\begin{table}[t]
\caption{Speech naturalness Mean Opinion Score (MOS) with 95\% confidence intervals across different language and systems.}
\vspace{-0.1in}
\centering
\begin{tabular}{c|c|c|c}
\toprule
\tbh{Languages} & EN & CN & ES \\
\midrule
\tbh{Monolingual C2A} & 4.24$\pm$0.12  & 3.48$\pm$0.11 & 4.21$\pm$0.11	\\
\tbh{Multilingual B2A} & 4.23$\pm$0.14  & 3.42$\pm$0.12 & 4.23$\pm$0.10\\
\bottomrule
\end{tabular}
\label{tbl:tts}
\vspace{-0.2in}
\end{table}
Table~\ref{tbl:tts} compares subjective naturalness MOS of the proposed model to the baseline using graphemes for English, Mandarin and Spanish respectively. 
Both results indicate that the proposed multilingual B2A model is comparable as the state-of-the-art monolingual model\footnote{MOS is worse than \cite{shen2018natural} because we have OOV in the test set.}. Moreover, we observed that the B2A model was able to read code-switching text. However, we don't have good metric to evaluate the quality of code-switching for TTS, e.g. the speech is fluent but the speaker is changed for different language. Future work may explore how to evaluate TTS on code-switching scenario and how to disentangle language and speaker given more training data.
\section{Conclusions}
\label{sec:concl}

In this paper, we investigated the use of Unicode bytes as a new language representation for both ASR and TTS. We proposed Audio-to-Byte (A2B) and Byte-to-Audio (B2A) as multilingual ASR and TTS end-to-end models. The use of bytes allows us to build a single model for many languages without modifying the model structure for new ones.
This brings representation sharing across graphemes, and is crucial for languages with large grapheme vocabularies, especially in multilingual processing. Our experiments show that byte models outperform grapheme models in both multilingual and monolingual models. Moreover, our multilingual A2B model outperforms our monolingual baselines by $4.4$\% relatively on average. The language independence of byte models provides a new perspective to the code-switching problem, where our multilingual A2B model achieves $38.6$\% relative improvement over our monolingual baselines. Finally, we also show our multilingual B2A models match the performance of our monolingual baselines in TTS.

\vfill\pagebreak

\bibliographystyle{IEEEbib}
\bibliography{refs}

\end{document}